\PassOptionsToPackage{super,sort&compress}{natbib}
\documentclass[aps,prb,preprint,groupedaddress]{revtex4-1}  
\usepackage{graphicx}
\usepackage{dcolumn}
\usepackage{bm}
\usepackage{amsmath}
\usepackage{amssymb}
\usepackage{mathtools}
\usepackage[utf8]{inputenc}
\usepackage[T1]{fontenc}
\usepackage[version=3]{mhchem}
\usepackage[colorlinks, allcolors=blue]{hyperref}
\hypersetup{pdftitle={Ionic Percolation Theory}, pdfauthor={Alexander Urban}}
\usepackage{microtype}
\newlength{\colwidth}
\begin{document}

\title{Modeling ionic transport and disorder in crystalline electrodes
  using percolation theory}

\author{Alexander Urban}
\email{a.urban@columbia.edu}
\affiliation{%
  Department of Chemical Engineering,
  Columbia University, USA.}
\date{\today}

\begin{abstract}
  \noindent
  Solid ionic conductors are essential components of batteries and fuel
  cells.
  In many cases, ionic conduction through crystalline materials with
  substitutional disorder can be modeled with atomic-scale lattice model
  percolation simulations.
  The ionic percolation theory reviewed in this chapter describes the
  percolation behavior of diffusion pathways, identifies the fraction of
  lattice sites that contributes to ionic conduction, and quantifies the
  tortuosity of diffusion networks.
  These quantities can be related to the bulk diffusivity and capacity
  of intercalation battery electrodes.
  We discuss applications to lithium-ion battery cathodes in the
  disordered rocksalt and related crystal structures, showing how the
  diffusion pathways and their tortuosity vary with the Li content
  short-range order.
  All examples are based on the free and open-source \emph{dribble}
  simulation software.
\end{abstract}

\maketitle

\section*{Introduction}

\noindent
Crystalline ionic conductors are essential for a variety of critical
technologies:
Lithium-ion (and other alkali ion) batteries require electrode materials
that are mixed ionic and electronic conductors,\cite{cr104-2004-4271}
and the ionic conductivity strongly affects the capacity and the rate
capability of the battery.\cite{acr46-2013-1216}
Ceramic ionic conductors are used as solid electrolyte in solid-state
batteries, which are promising candidates for the next generation of
Li-ion batteries,\cite{cr116-2016-140, ne1-2016-2016141} and developing
stable solid electrolytes with high ionic conductivity is one of the key
challenges.
The solid electrolyte materials in solid-oxide fuel and electrolyzer
cells (SOCs) are oxide-ion (\ce{O^2-}), proton (\ce{H+}), or hydride
(\ce{H-}) ion conductors,\cite{csr39-2010-4370} and the operation
temperature required by SOCs is, to a large extent, determined by the
ionic conductivity of the solid electrolyte.\cite{jps339-2017-103}
Ionic conductors have also been proposed as components of
memristors,\cite{n453-2008-80} non-linear electrical components that
could provide new circuit functions.

Atomic-scale simulations have been instrumental in obtaining mechanistic
insight into ionic diffusion on the atomic scale.\cite{ees4-2011-2774,
  acr46-2013-1216}
However, since such simulations rely on idealized structure models with
small periodic units, they cannot provide information about the nature
of extended diffusion pathways in materials with substitutional
disorder, a common phenomenon in ionic conductors.
Here we discuss how lattice model simulations can complement atomistic
simulations and experimental characterization to identify diffusion
pathways in disordered crystalline ionic conductors.

In the next section, we will briefly review select previous work on
simulating ionic percolation in crystalline materials.
This is followed by a description of the ionic percolation methodology
and its implementation in the free and open-source \textit{Dribble}
software.
We will then discuss examples of applications in the area of battery
materials before considering the strengths and limitations of the
methodology in a final discussion section.

\section{Background}

\subsection{Ionic percolation in crystalline solids}

\begin{figure}
  \centering
  \includegraphics[width=0.8\colwidth]{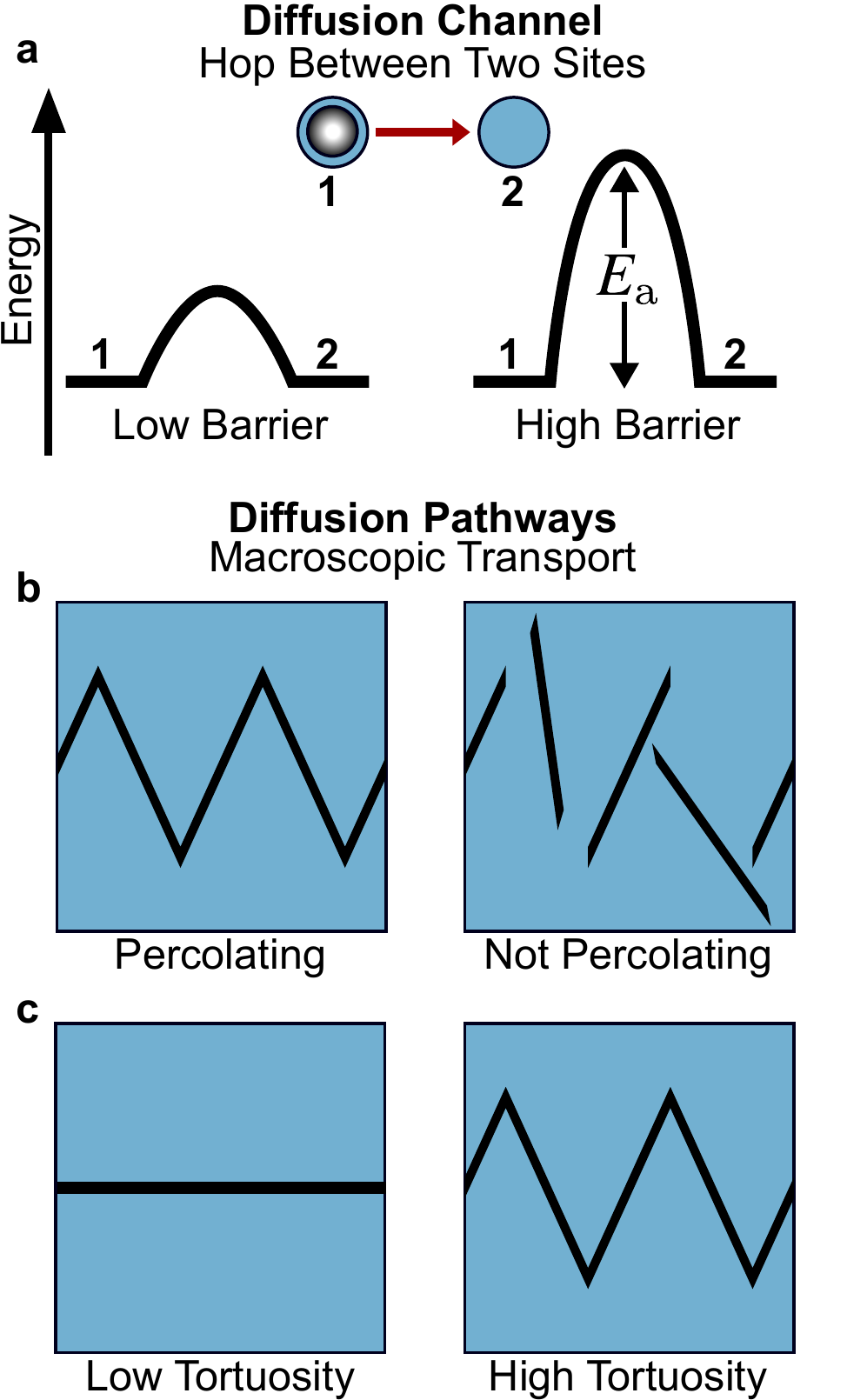}
  \caption{\label{fig:barrier-percolation-tortuosity}%
    \textbf{Schematic of three ingredients required for good ionic
      conduction.}  \textbf{a}, On the atomic scale, diffusion requires
    active (low-barrier) diffusion channels so that the diffusing
    species can hop from one site to another.  \textbf{b}, The active
    diffusion channels have to connect to percolating diffusion pathways
    so that the ions can migrate through the entire material.
    \textbf{c}, For efficient conduction (high diffusivity), the
    tortuosity of the diffusion pathways should be low.}
\end{figure}

\noindent
A crystalline ionic conductor is a solid material through which a
specific type of ion can diffuse over macroscopic length scales.
For example, Li ions in a Li-ion battery (LIB) diffuse through the bulk
of the electrode particles during the charging and discharging of the
battery.
Two requirements have to be met so that macroscopic ionic transport is
possible in a given material:
\begin{itemize}
\item[(i)] There have to exist diffusion channels that are
  \textit{active} at operation conditions, \emph{i.e.}, at room
  temperature and under realistic applied potentials in the case of
  batteries (\textbf{Fig.~\ref{fig:barrier-percolation-tortuosity}a}),
  and
\item[(ii)] These channels need to be interconnected in
  \textit{percolating} diffusion pathways that span the material
  (\textbf{Fig.~\ref{fig:barrier-percolation-tortuosity}b}).
\end{itemize}

In this picture, a diffusion channel is defined as the minimum-energy
pathway connecting two lattice sites.
Whether a diffusion channel is active depends on the activation energy
that needs to be overcome when an ion migrates through the channel.
For given conditions (temperature, chemical potentials, electric field,
\emph{etc.}), the activation energy determines the rate with which ions
hop from one of the lattice sites to the other.
The magnitude of an acceptable activation energy depends on the
application.
For example, an activation barrier of $\leq$300~meV is typical for Li
diffusion in LIB cathodes.\cite{ncm2-2016-16002}
The activation energy for \ce{O^2-} hopping in SOCs is much greater,
which is the reason why SOCs need to be operated at high temperatures.
First principles calculations are routinely used for predicting the
activation energy associated with a given diffusion
channel.\cite{nm14-2015-1026, nc7-2016-11009}

Even when active diffusion channels are present, a material might still
not be an ionic conductor if the channels do not form a network across
the entire material.
The conditions for the formation of such \emph{percolating} diffusion
pathways is the focus of the present chapter.
In addition to percolation, we will also consider the \emph{tortuosity}
of diffusion pathways which is a measure of how direct the pathways are
(\textbf{Fig.~\ref{fig:barrier-percolation-tortuosity}c}).

In an ordered crystal structure it is often obvious from visual
inspection if the active diffusion channels form percolating diffusion
pathways.
For example, a two-dimensional network of percolating Li diffusion
pathways is available in layered \ce{LiCoO2}\cite{esl3-2000-301} and
graphite,\cite{jpcl1-2010-1176} and one-dimensional pathways exist in
\ce{LiFePO4}.\cite{esl7-2004-30, com17-2005-5085}
A negative example is \ce{Na3GaV(PO4)2F3}, in which fast isolated Na
diffusion channels exist but do not form percolating
pathways.\cite{cm27-2015-6008}

However, many crystalline ionic conductors exhibit substitutional
disorder, \emph{i.e.}, sites that can be occupied by different atomic
species.
In such disordered materials, the percolation properties might no longer
be obvious, since the activation energy of a diffusion channel can
depend sensitively on the neighboring atomic species.
Important examples of such disordered ion conductors are
cation-disordered cathode materials for LIBs\cite{sci343-2014-519,
  pnsa112-2015-7650, aem5-2015-1401814, n556-2018-185} and common solid
electrolytes for solid oxide fuel and electrolyzer cells (SOCs), such as
yttria-stabilized zirconia\cite{cm23-2011-1365} and doped
ceria.\cite{prb77-2008-024108}
Note that long-range disorder does not equal a fully random distribution
on shorter length scales, and short-range order can significantly affect
ion transport.\cite{nc10-2019-592}

\subsection{Diffusion mechanism and diffusion channels}

\begin{figure}
  \centering
  \includegraphics[width=1.0\colwidth]{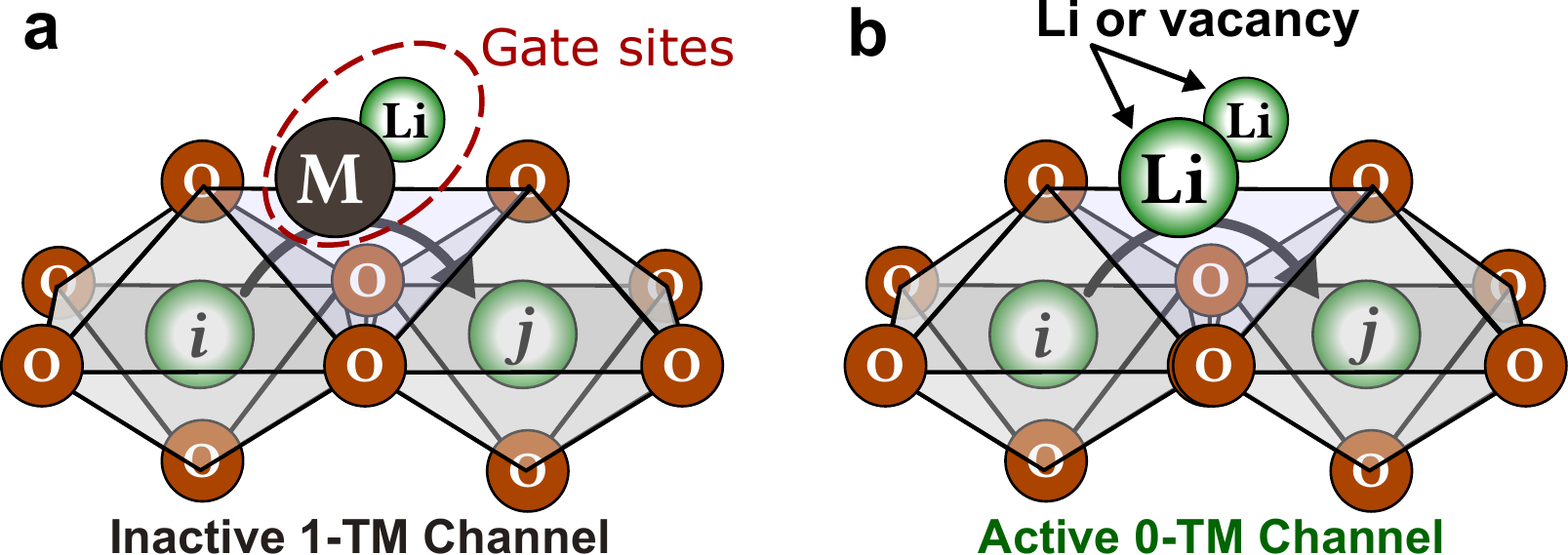}
  \caption{\label{fig:channels}%
    \textbf{Li diffusion channels in disordered rocksalt-type oxides.}%
    Li diffusion between two octahedral sites proceeds via a tetrahedral
    activated state with neighboring \emph{gate sites}.
    Diffusion along one or more neighboring transition metal (TM) ion(s)
    is associated with a high activation barrier.
    The activation barrier is typically the lowest for diffusion
    through tetrahedral sites coordinated only by Li or vacancies
    (active 0-TM channels).
    In disordered structures 1-TM and 2-TM channels are typically inactive.
    The transition metal ion is labeled with an M in the figure.  }
\end{figure}

\noindent
In the following, we consider rocksalt-type LIB cathodes as a concrete
example of a class of crystalline ionic conductors with substitutional
disorder.
Nevertheless, the discussed concepts are transferable to other crystal
structures and materials.

Li diffusion in rocksalt-type Li transition-metal (TM) oxides has been
extensively studied with experiments and simulations, as reviewed by
Van~der~Ven et~al.\cite{acr46-2013-1216}
In these materials, the cation (Li and TM) and anion (oxygen) sites are
on separate face-centered cubic sublattices.
The cation sites are octahedral, \emph{i.e.}, each site is coordinated
by six oxygen ions, and Li diffusion between two octahedral (\emph{o})
sites takes place via a tetrahedral (\emph{t}) activated state in the
presence of a second vacancy (\emph{o--t--o} di-vacancy
mechanism).\cite{esl3-2000-301, acr46-2013-1216}
The activation barrier for Li diffusion depends on the species on the
two \emph{gate sites} that coordinate the tetrahedral site along the
o--t--o diffusion channel (\textbf{Fig.~\ref{fig:channels}}), and
because of electrostatic repulsion, the barrier is generally greater if
one or both of the gate sites are occupied by TM ions (1-TM and 2-TM
channels) and lower for Li ions or vacancies (0-TM
channels).\cite{acr46-2013-1216, sci343-2014-519}
In ordered crystal structures such as the layered \ce{LiCoO2} structure
the diffusion channels are within the Li layer and are typically larger
than in cation-disordered structures.\cite{sci343-2014-519}
As a consequence, 1-TM diffusion channels are typically active in the
layered structure but are inactive in cation-disordered
structures.\cite{sci343-2014-519}
Therefore, 0-TM channels are the most important diffusion channel in
cation-disordered rocksalt-type materials, which can be good Li-ion
conductors if their 0-TM channels form percolating diffusion pathways.

\section{Method}

\subsection{Lattice percolation theory}
\label{sec:lattice-percolation-theory}

\noindent
Percolation theory, and specifically the percolation of sites and bonds
on lattices, is a standard problem in statistical and computational
physics,\cite{Stauffer1994, LandauBinder2009} and has a long history in
the simulation of transport phenomena.\cite{rmp64-1992-961}
The general \emph{site percolation} problem is as follows:
Given an infinite lattice with randomly vacant or occupied sites, what
concentration of occupied sites is required so that nearest-neighbor
bonds between occupied sites span infinitely across the lattice,
\emph{i.e.}, become percolating.
On an infinite lattice, there is exactly one critical concentration of
occupied sites $x_{c}$, the \emph{percolation threshold}, below which
the system is never percolating and above which it is always
percolating.
The percolation probability $p(x)$ is thus a step function
\begin{align}
  p(x) = \begin{cases}
    0 & \text{if}\; x < x_{c} \\
    1 & \text{if}\; x > x_{c}
  \end{cases}
  \label{eq:percolation-threshold}
  \quad .
\end{align}

For many regular two-dimensional lattices the percolation threshold can
be calculated analytically, but for three-dimensional lattices numerical
Monte Carlo (MC) simulations are needed to obtain approximate
values.\cite{prl85-2000-4104, LandauBinder2009}
Practical simulations employ finite lattices or periodic boundary
conditions, resulting in finite-size errors that lead to deviations from
the ideal case.
For finite lattices, the percolation probability is a sigmoidal function
that approaches the step function of
equation~\eqref{eq:percolation-threshold} with increasing lattice (or
unit cell) dimension.
The percolation threshold can be approximated by the inflection point of
the sigmoidal function but needs to be converged with respect to the
cell size.

\subsection{Application of lattice percolation theory to ionic transport}
\label{sec:application-to-ionic-transport}

\noindent
Ionic percolation simulations have been instrumental in understanding
under which conditions cation-disordered LIB cathodes become Li
conducting and allow for the reversible extraction of large Li
capacities.\cite{sci343-2014-519, aem-2014-1400478, nc10-2019-592}
Percolation simulations have been used to elucidate the impact of
fluorine doping on Li-ion conduction,\cite{aem10-2020-1903240} and they
have led to the discovery of highly energy-dense partially disordered
spinel LIB cathodes.\cite{ne5-2020-213}
Apart from applications to LIBs, the method has also been used to
investigate the effect of inversion on Mg-ion conduction in spinel-type
materials\cite{cm29-2017-7918} and Na-ion conduction in Na-ion battery
cathodes.\cite{mtc28-2023-101368}

The formation of percolating diffusion pathways in ion conductors can be
described by site percolation if the lattice decoration uniquely
determines which neighboring sites are connected by active diffusion
channels.
The precise relationship between site occupancies and active/inactive
diffusion channels is generally more complex than the nearest-neighbor
criterion of the conventional site percolation problem of
section~\ref{sec:lattice-percolation-theory}.
Defining criteria for active diffusion channels requires prior knowledge
of the relationship between lattice occupancies and diffusion barriers,
such as the 0-TM channels in cation-disordered rocksalts discussed
above.
Criteria for active diffusion channels can often be defined in terms of
\emph{bond} or \emph{site rules}.

\emph{Bond rules} are a set of criteria that determine if two
neighboring sites $i$ and $j$ are connected by an active diffusion
channel.
In the simplest case, such bond rules can depend only on the common
neighboring sites of sites $i$ and $j$.
For example, a 0-TM channel exists between sites $i$ and $j$ if (i)~at
least two sites $k$ and $l$ that are nearest neighbors of both $i$ and
$j$ are also Li sites; and (ii)~$k$ and $l$ are themselves nearest
neighbors.
This definition identifies two gate sites
(\textbf{Fig.~\ref{fig:channels}}) and ensures that they are both Li
sites.

\emph{Site rules} define neighbor shell decorations for which a site of
a specific sublattice may become part of a diffusion pathway.
For example, 0-TM channels can be defined in terms of a site rule by
explicitly including the sublattice of the tetrahedral transition states
and requiring that only those tetrahedral sites that are surrounded by
4~octahedral Li sites are considered.
For rocksalt-type structures, this site rule definition might appear
less intuitive than the above bond rule, since it involves two types of
sites: tetrahedral and octahedral sites.
However, it is the more natural framework for partially inverted spinel
structures, where vacancies have to be explicitly
considered.\cite{cm29-2017-7918}
Note that the octahedral cation sites of the rocksalt structure and the
tetrahedral intermediates together define the fluorite (\ce{CaF2})
structure.

\subsection{Detecting percolation in simulations}

\noindent
The most fundamental question to address with percolation theory is
whether a given crystal structure with integer site occupancies
(\emph{i.e.}, without disorder) is percolating.
For large unit cells and complex atomic ordering, for example if atomic
configurations were themselves obtained from MC
simulations,\cite{cm29-2017-7840} it might not be immediately obvious if
percolating diffusion pathways exist.

\begin{figure}
  \centering
  \includegraphics[width=\colwidth]{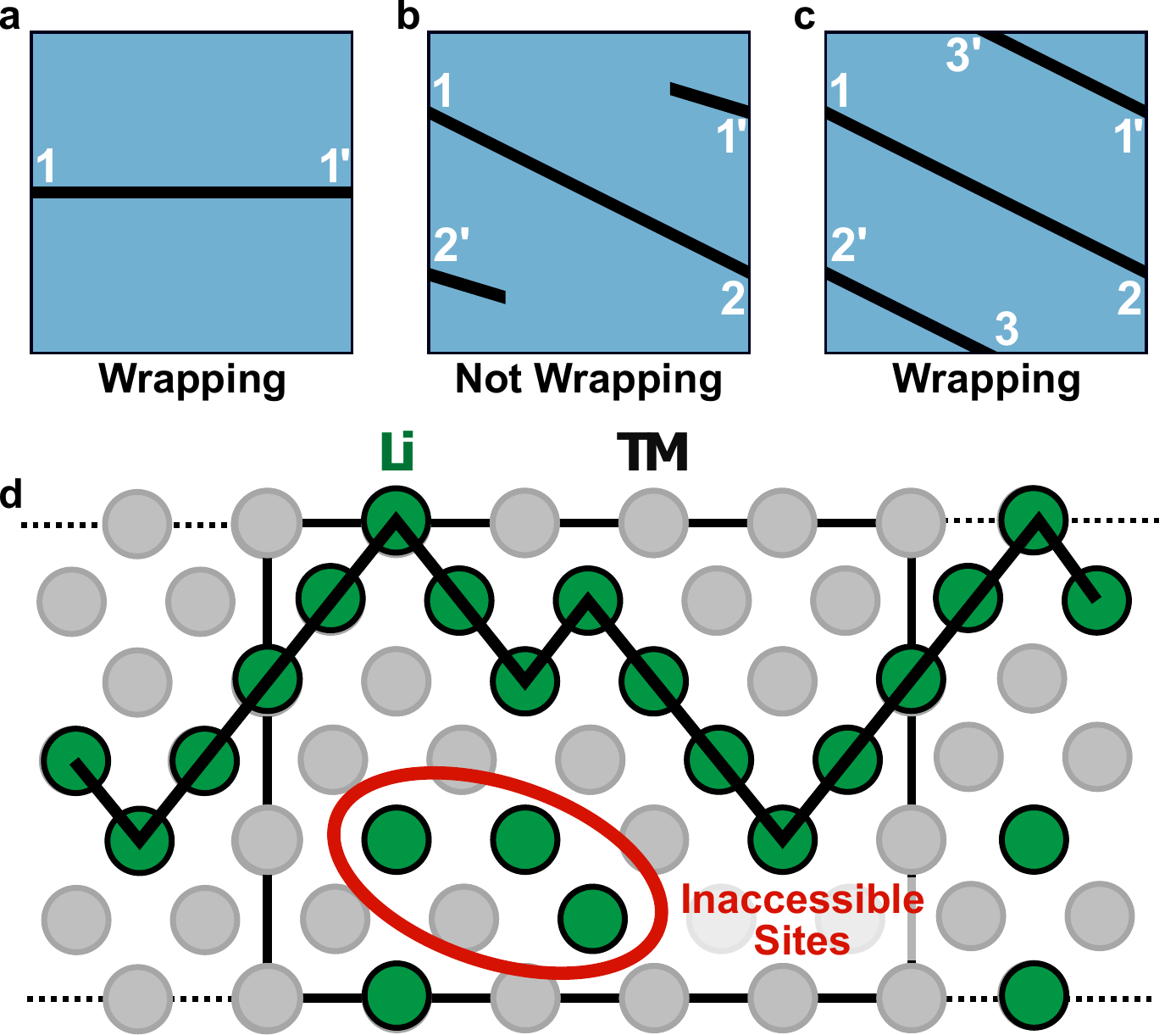}
  \caption{\label{fig:wrapping}%
    \textbf{Schematics of unit cells with periodically wrapping and
      non-wrapping diffusion pathways.}
    \textbf{a}, A periodically wrapping diffusion pathway that
    connects two periodic images $1$ and $1'$ of the same site.
    \textbf{b}, Non-wrapping pathway between two different sites $1$ and
    $2$.  The pathway $1\longrightarrow{}2$ does not extend to the
    periodic image of site $1$.
    \textbf{c}, Periodically wrapping pathway that involves three
    segments $1\longrightarrow{}2$, $2'\longrightarrow{}3$, and
    $3'\longrightarrow{}1'$ crossing the periodic boundary conditions.
    \textbf{d}, Schematic of a wrapping nearest-neighbor pathway (black
    line) and inaccessible sites that are not connected to the
    percolating pathway (encircled in red).  Occupied sites
    corresponding to active diffusion channels are green, and vacant
    sites are gray.}
\end{figure}

On a finite lattice with periodic boundary conditions, percolation is
equivalent to periodic wrapping:
If a diffusion pathway exists that connects one site in the structure
with any periodic image of the same site, then the structure is
percolating (\textbf{Fig.~\ref{fig:wrapping}a-c}).
Note that a periodically wrapping diffusion pathway can cross the
boundaries of the crystal unit cell multiple times
(\textbf{Fig.~\ref{fig:wrapping}c}).

Efficient algorithms to detect periodic wrapping have previously been
developed, such as the method by Newman and Ziff,\cite{prl85-2000-4104}
which incrementally grows interconnected domains of sites and detects
when a domain connects to a periodic image of an already connected site.

\subsection{Accessible sites}
\label{sec:accessible-sites}

Even though a given structure is percolating, not all sites of the
diffusing species are necessarily connected to a percolating diffusion
pathway (Figure~\ref{fig:wrapping}d).
This means, there can exist isolated domains of sites in a percolating
structure that are occupied by the diffusing species but do not
contribute to ionic conduction.
We define the fraction of \emph{accessible sites} as the number of sites
$N_{\textup{p}}$ that are connected to percolating diffusion pathways
divided by the total number of sites of the diffusing
species ($N_{\textup{tot}}$)
\begin{align}
  x_{\textup{acc}} = \frac{N_{\textup{p}}}{N_{\textup{tot}}}
  \quad .
  \label{eq:fraction-of-accessible-sites}
\end{align}

The fraction of accessible sites is important, for example, in LIB
cathodes where the amount of extractable Li determines the
\emph{capacity}, and inaccessible sites thus do not contribute to the
practical capacity of the cathode and are essentially unused
material.\cite{aem-2014-1400478}
The capacity is either given relative to the composition-specific mass
or volume of the material or it is normalized to a formula unit.
For example, the capacity in terms of Li atoms per \ce{Li_xM_{2-x}O2}
formula unit is then $C = x_{\textup{acc}} x$.

\subsection{Tortuosity}
\label{sec:tortuosity}

\noindent
Two percolating structures with the same amount of accessible sites are
not necessarily equally good ionic conductors.
Another factor that contributes to the overall ionic conductivity is the
\emph{tortuosity} of the diffusion pathways, \emph{i.e.}, the length of
the diffusion pathway $L$ relative to the net distance $D$ that an ion
travels
\begin{align}
  \tau = \frac{L}{D}
  \quad.
\end{align}
The tortuosity thus corresponds to a detour that the diffusing ion
takes, and one can expect that a high tortuosity is detrimental for
ionic conduction.
For example, in the case of mass transport in porous media, the
tortuosity of the medium gives rise to a scaling factor of the diffusion
constant which has been extensively studied.\cite{ces62-2007-3748}
The effect of tortuosity on ionic transport in composite materials such
as the electrodes in LIBs\cite{jps188-2009-592} and
SOCs\cite{nm5-2006-541} is also an important performance parameter.

In lattice model percolation simulations, the tortuosity of a given
structure can be determined by calculating the shortest diffusion
pathway between each site in a percolating domain and all of its
periodic images that are part of the same domain.
The length of the shortest pathway $L_{i}$ divided by the distance
$D_{ii'}$ between site $i$ and its periodic image $i'$ is then the
tortuosity for site $i$.
We define the tortuosity of a structure for ionic conductivity as the
mean tortuosity of all sites
\begin{align}
  \tau_{\textup{ion}}
  = \sum_{i}^{\textup{sites}} \frac{L_{i}}{D_{ii'}}
  = \left\langle \frac{L_{i}}{D_{ii'}} \right\rangle
  \quad .
  \label{eq:ionic-tortuosity}
\end{align}

\subsection{Lattice percolation simulations with \emph{Dribble}}

\noindent
The free and open-source simulation software \emph{Dribble}
(\url{https://github.com/atomisticnet/dribble}) implements an MC method
similar to that by Newman and Ziff\cite{prl85-2000-4104} for the
calculation of all of the quantities discussed above, the percolation
threshold, the fraction of accessible sites, and the ionic tortuosity.

Apart from analyzing static structures, the response of the percolation
properties to composition changes and/or to changes in the degree of
substitutional disorder is often of interest.
For example, 0-TM channels are a local lithium-rich environment, and
therefore their concentration increases with the Li content,
\emph{i.e.}, we can identify a percolation threshold for 0-TM channels
in terms of the Li content.
For fully Li/M disordered rocksalt-type structures with composition
\ce{Li_xM_{1-x}O2}, this percolation threshold lies at around
$x_{c}\approx{}1.09$.\cite{sci343-2014-519}
Remnants of an ordered structure, such as short-range order similar to
the layered or spinel structure, additionally alter the percolation
threshold.\cite{aem-2014-1400478, nc10-2019-592}

\emph{Dribble} therefore also provides MC algorithms for the scanning of
compositions and for the sampling of disordered structures.
Substitutional disorder is simulated by repeating the analysis for a
user-specified number of randomly generated lattice decorations.
Here, the composition of each sublattice can be defined independently.
For example, in a fully disordered \ce{(Li, M)2O2}, each cation site is
occupied by either Li or M with equal probability.
To scan a composition range \ce{M2O2}--\ce{Li2O2}, the simulation is
initialized with composition \ce{M2O2} followed by the iterative
flipping of randomly selected M sites to Li until the composition
\ce{Li2O2} has been reached.
The convergence of the wrapping probability with MC sampling can be
accelerated by evaluating the convolution of the observed wrapping
probability with a binomial distribution.\cite{prl85-2000-4104}

Random sampling becomes more accurate with increasing size of the
simulation cell, and care has to be taken to converge computed
quantities with respect to the cell size.
Owing to finite-size effects, as mentioned above, the periodic wrapping
probability resembles a sigmoidal function for small cell sizes and
approaches the ideal step function of
equation~\eqref{eq:percolation-threshold} with increasing cell
size.\cite{LandauBinder2009}
On the other hand, the computational cost also increases with the cell
size so that the smallest cell should be determined that gives the
desired precision.

All calculations of percolation thresholds, wrapping probabilities, and
accessible sites reported in this chapter were done over 500~MC steps.
Tortuosity calculations generally were done for 50~MC steps because of
their greater computational cost.

\section{Examples of lattice percolation simulations}

To illustrate the concepts outlined above, this section discusses two
concrete LIB cathode materials:
fully disordered rocksalts \ce{(Li_{x}, M_{1-x})2O2} and orthorhombic
\ce{LiMnO2} with varying degree of Li~$\leftrightarrow{}$Mn mixing.
For both systems, 0-TM channels were taken to be the only active
diffusion channels, and the \emph{bond rule} definition described in
section~\ref{sec:application-to-ionic-transport} was used for channel
detection.

\begin{figure*}
  \centering
  \includegraphics[width=1.8\colwidth]{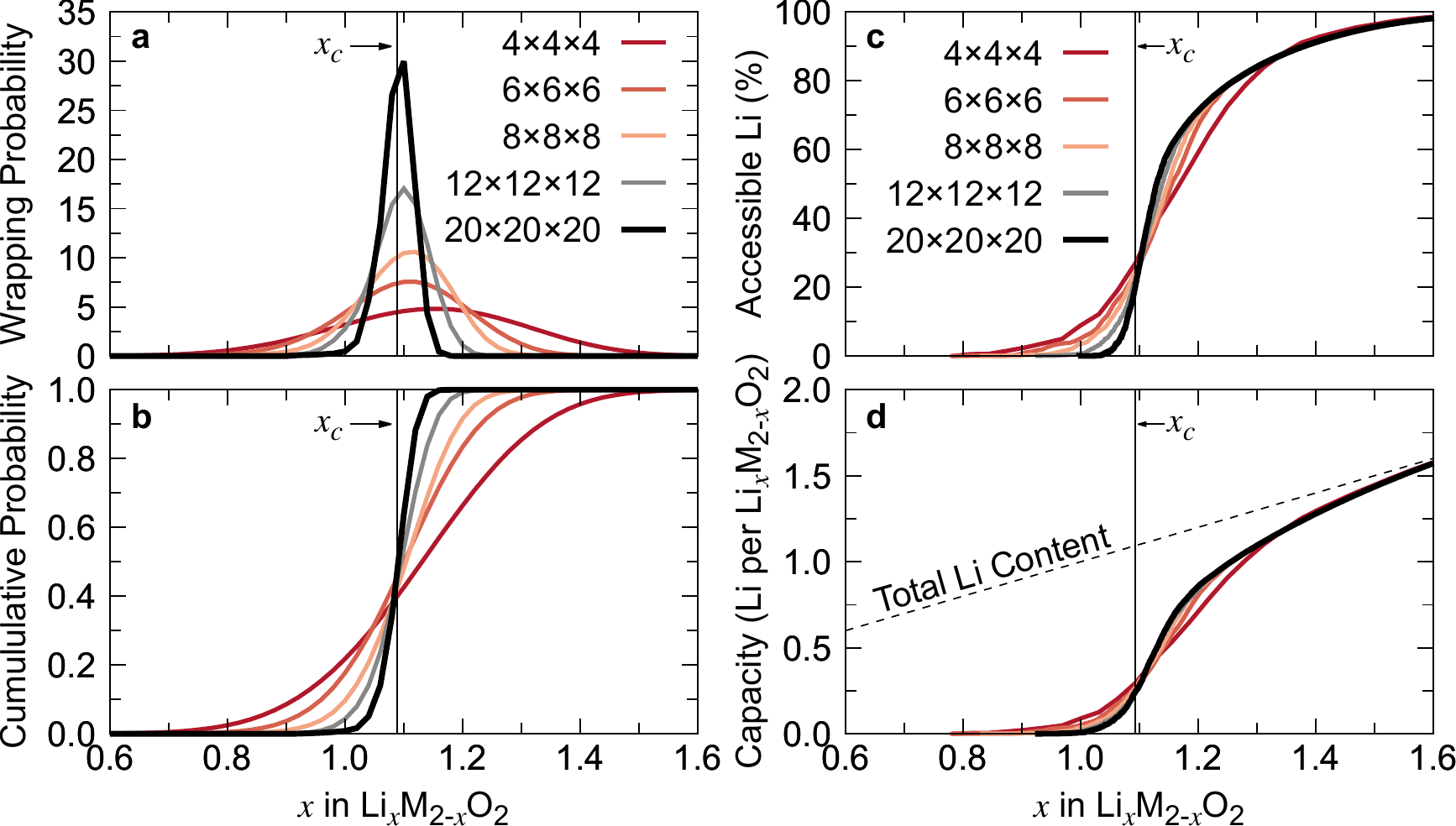}
  \caption{\label{fig:rocksalt}%
    \textbf{0-TM wrapping probability and accessible capacity.}
    Convergence of the \textbf{a-b}, periodic wrapping probability
    (binomial convolution) and \textbf{c-d}, accessible sites with the
    cell size.
    \textbf{a} shows the \emph{wrapping probability}, \emph{i.e.}, the
    probability of a given Li content to be the lowest Li concentration
    for which periodic wrapping occurs in the Monte Carlo simulation.
    This quantity approaches a delta function at the percolation
    threshold ($x_{c}\approx{}1.09$) with increasing cell size.
    The corresponding cumulative wrapping probability is shown in
    \textbf{b} and approaches a step function.
    \textbf{c} shows the percentage of Li sites within the structure
    that are connected to percolating diffusion pathways.
    This quantity can be converted to a capacity in terms of extractable
    Li content per \ce{Li_xM_{2-x}O2} formula unit which is shown in
    \textbf{d}.
    All cell dimensions are given relative to the primitive rocksalt
    unit cell with one cation and one anion site, \emph{e.g.}, the
    $10\times{}10\times{}10$ supercell contains 1\,000 cation sites.}
\end{figure*}

\subsection{Properties of fully disordered rocksalts}


\noindent
Fully disordered crystal structures are most similar to the conventional
lattice percolation problem of statistical physics.
The first system that we consider here is therefore the ideal \ce{(Li,
  M)_2O2} disordered rocksalt structure in which \ce{Li} and \ce{M} atoms
share the same sublattice and are statistically distributed.
\textbf{Figure~\ref{fig:rocksalt}} shows the convergence of the periodic
wrapping probability and accessible sites in a rocksalt structure with
the size of the simulations cell.
The periodic wrapping probability plotted in
\textbf{Figure~\ref{fig:rocksalt}a} is the probability distribution
function of the lowest Li content for which periodic wrapping was
detected in the MC simulation.
This quantity is determined in MC simulations by initially decorating
the cation sublattice only with M atoms and incrementally increasing the
Li content until periodically wrapping diffusion pathways are observed.

As discussed in section~\ref{sec:lattice-percolation-theory}, the
percolation probability is a step function for an infinitely extended
lattice.
The probability of the lowest percolating concentration thus is a
(Dirac) delta function centered at the percolation threshold $x_{c}$
\begin{align}
  p_{1}(x)
  = \delta(x-x_{c})
  = \begin{cases}
      \infty & \text{for}\; x = x_{c} \\
      0 & \text{otherwise}
    \end{cases}
  \quad .
\end{align}
As seen in \textbf{Figure~\ref{fig:rocksalt}a}, this delta function is
broadened to continuous distributions for finite cell sizes.
Importantly, the distributions are not exactly centered around the true
percolation threshold $x_{c}$, and using the peak of a distribution as
an approximation for $x_{c}$ can lead to a substantial overestimation of
the percolation threshold for small cell sizes.

\textbf{Figure~\ref{fig:rocksalt}b} shows the corresponding cumulative
distribution function of the periodic wrapping probability (\emph{i.e.},
the integral of \textbf{Figure~\ref{fig:rocksalt}b}), which approaches
the percolation probability $p(x)$ of
equation~\eqref{eq:percolation-threshold} with increasing cell size.
The percolation threshold is a Li content of $x_{c}\approx{}1.09$,
\emph{i.e.}, around 10\% Li excess compared to the stoichiometric
\ce{LiMO2} composition, which is consistent with our previous
report.\cite{aem-2014-1400478}

As seen in \textbf{Fig.~\ref{fig:rocksalt}c}, the fraction of accessible
sites as function of the Li content also needs to be converged with
respect to the simulation cell.
Although, it converges rapidly for Li contents well above the
percolation threshold, convergence near $x_{c}$ requires large cell
sizes.
Per definition, on an ideal infinite lattice no sites are accessible
below the percolation threshold, and thus the predicted amount of
accessible sites for $x<x_{c}$ is a finite-size artifact that must
vanish for increasingly larger supercells.
As discussed in section~\ref{sec:accessible-sites}, the fraction of
accessible Li sites can be converted into the expected \emph{capacity}
of an LIB cathode and is therefore a key quantity for composition
optimization (\textbf{Fig.~\ref{fig:rocksalt}d}).

\begin{figure*}
  \centering
  \includegraphics[width=1.8\colwidth]{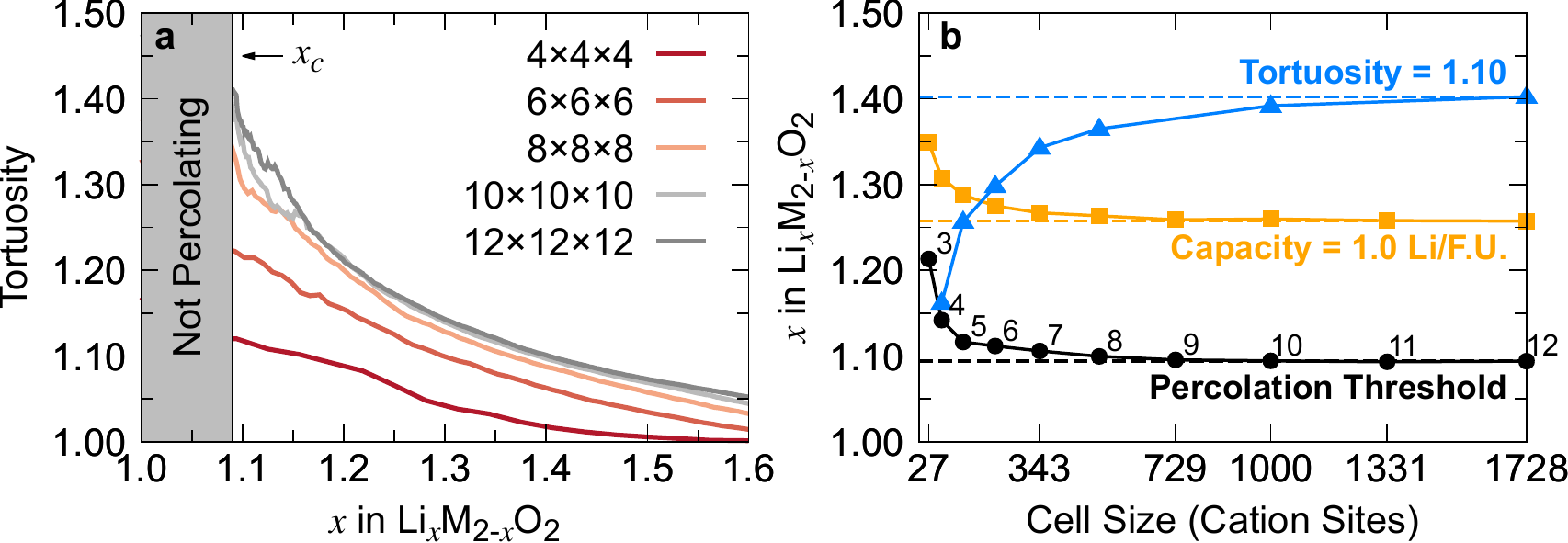}
  \caption{\label{fig:rocksalt-2}%
    \textbf{Tortuosity and cell size dependence of the Monte Carlo
      results.}
    \textbf{a}, Tortuosity of percolating diffusion pathways in
    cation-disordered rocksalt structures with increasing supercell
    size.
    \textbf{b}, Convergence of the percolation threshold
    ($x_{c}\approx{}1.09$), the composition that yields a capacity of
    1~Li atom per \ce{Li_xM_{2-x}O2} formula unit ($x\approx{}1.26$),
    and the composition with a tortuosity of~1.10 ($x\approx{}1.40$)
    with the cell size.
    Cell dimensions are given in terms of the primitive rocksalt unit
    cell with one cation and one anion site.}
\end{figure*}
\textbf{Figure~\ref{fig:rocksalt-2}a} shows the tortuosity
$\tau_{\textup{ion}}$ of 0-TM diffusion pathways in increasingly larger
cells of the disordered rocksalt structure.
The tortuosity can only be evaluated for percolating structures, so that
only Li contents above the percolation threshold are meaningful.
Note that tortuosity calculations using the definition of
$t_{\textup{ion}}$ of equation~\eqref{eq:ionic-tortuosity} are
computationally demanding for large cell sizes, as they involve the
calculation of the length of all possible percolating diffusion pathways
(see section~\ref{sec:tortuosity}).
As seen in the figure, the tortuosity of the $8\times{}8\times{}8$ cell
is essentially converged, so that no data for cell sizes beyond the
$12\times{}12\times{}12$ supercell (1\,728~sites) are shown.
Right at the percolation threshold $x_{c}$, the tortuosity is greater
than $\tau=1.40$, \emph{i.e.}, 0-TM diffusion pathways are on average
40\% longer than the distance that they span.
Realistic compositions for high-capacity cathodes with $x\approx{}1.2$
still have an average tortuosity of $\sim$20\%.

The cell-size convergence tests for the percolation threshold, a
realistic target capacity, and a realistic target tortuosity are
summarized in \textbf{Figure~\ref{fig:rocksalt-2}b}.
As seen in the figure, even though the wrapping probability and
accessible Li sites converge slowly near the percolation threshold, the
threshold and the capacity are already converged to within $\pm{}0.01$
in terms of the Li content for an $8\times{}8\times{}8$ supercell.
Since, the 0-TM capacity is a lower bound of the capacity and not an
absolute prediction, a greater precision should not be needed.
Because of the slow decay of the tortuosity with the Li content, the
error in the Li content for a given tortuosity value is slightly higher
($\pm{}0.04$) but probably also within an acceptable tolerance.
Thus, for practical applications cell sizes with around 500~sites
($8\times{}8\times{}8=512$~cation sites) yield quantitative results.

\subsection{Li percolation in orthorhombic \ce{LiMnO2}}

\begin{figure}
  \centering
  \includegraphics[width=0.8\colwidth]{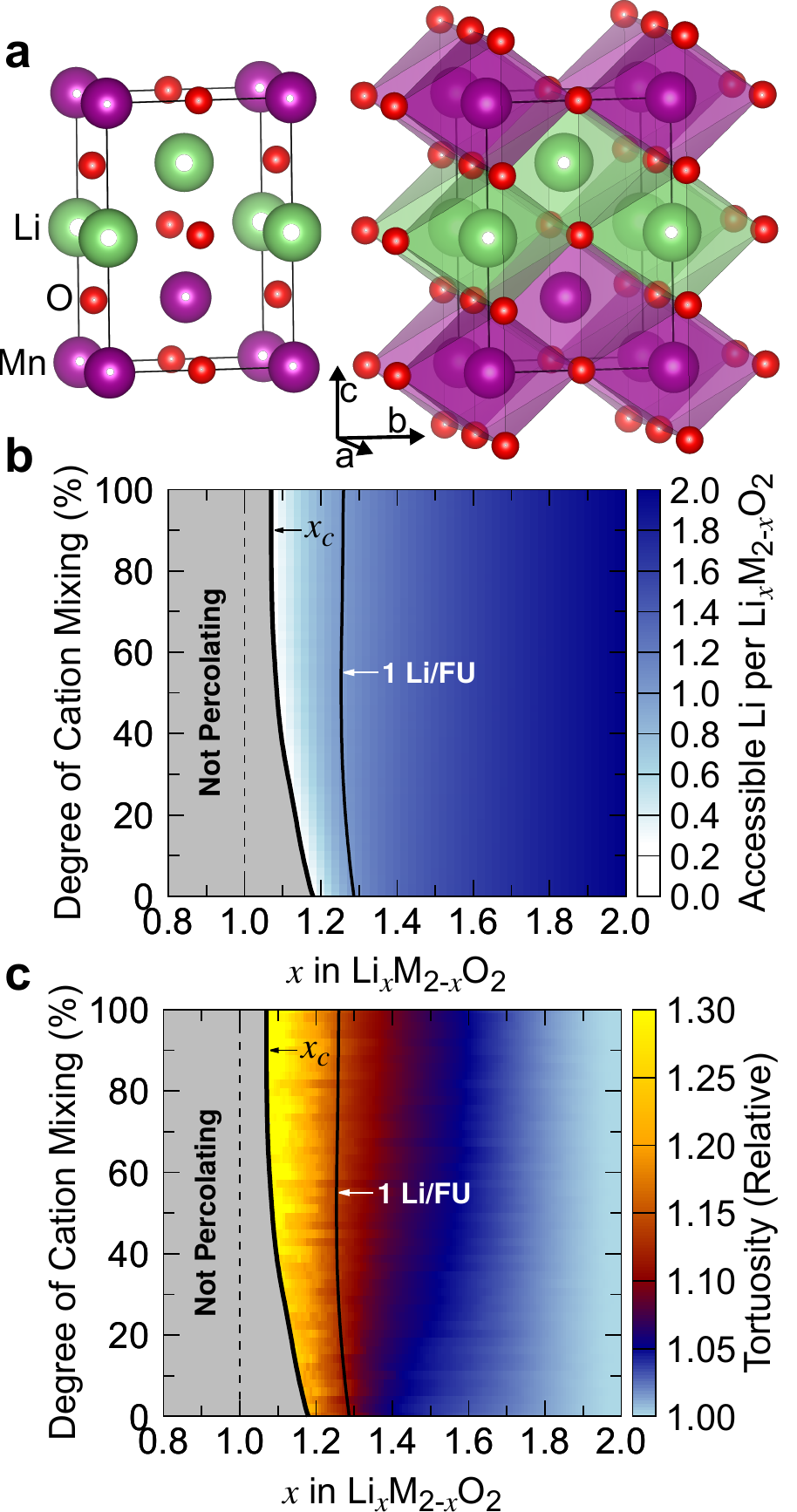}
  \caption{\label{fig:o-LiMnO2}%
    \textbf{Percolation and tortuosity in \ce{o-LiMnO2}.}
    \textbf{a}, Unit cell of the ideal orthorhombic \ce{LiMnO2} crystal
    structure (space group $Pmmn$) corresponding to two formula units,
    visualized with VESTA.\cite{jac41-2008-653,
      jac44-2011-1272}
    The cation sites are octahedral, \emph{i.e.}, coordinated by each six
    oxygen atoms.
    Note that the actual structure of \ce{o-LiMnO2} is distorted from
    the ideal structure shown in the figure by Jahn-Teller
    distortions.\cite{jpcs48-1987-97}
    \textbf{b}, 0-TM capacity as function of the Li content and
    the degree of cation mixing.
    The ordered \ce{o-LiMnO2} structure corresponds to 0\% cation
    mixing, whereas 100\% cation mixing corresponds to the fully
    disordered rocksalt structure (results for which are shown in
    figure~\ref{fig:rocksalt}).
    A $6\times{}6\times{}6$ supercell (864~cation sites) of the
    \ce{o-LiMnO2} unit cell of \textbf{a} was used for the calculations.
    \textbf{c}, Tortuosity as function of the Li content and the degree
    of cation mixing based on a $4\times{}4\times{}4$ supercell
    (256~cation sites).
    In \textbf{b} and \textbf{c}, the percolation threshold ($x_{c}$) is
    indicated by a thick black line, and a thin black line indicates the
    conditions at which the 0-TM capacity is 1~Li per formula unit.}
\end{figure}

\noindent
Statistical (random) disorder is an idealization, and in real materials
some degree of cation ordering is likely to be present.
It is therefore useful to inspect how the percolation behavior changes
when an ordered crystal structure gradually becomes disordered,
\emph{e.g.}, by mixing the Li and M sublattices in the case of LIB
cathode materials.
Such percolation maps for the gradual disordering of the layered
(\ce{\alpha-NaCoO2}) structure, the low-temperature \ce{Li2Co2O4} spinel
structure, and the \ce{\gamma-LiFeO2} structure have previously been
reported.\cite{aem-2014-1400478}
Motivated by the recent interest in Mn-based disordered rocksalts and
spinels,\cite{pnsa112-2015-7650, n556-2018-185, ne5-2020-213}
we consider here the orthorhombic \ce{LiMnO2} (\ce{o-LiMnO2}) crystal
structure (\textbf{Fig.~\ref{fig:o-LiMnO2}a}) which is the thermodynamic
ground-state crystal structure of \ce{LiMnO2} at ambient
conditions.\cite{jpcs48-1987-97}
Cation-disordered LIB cathode materials can exhibit short-range order
that is similar to the cation ordering in related ordered
oxides,\cite{nc10-2019-592} so compositions containing a significant
amount of \ce{Mn(III)} might exhibit short-range order that is related
to the cation ordering in \ce{o-LiMnO2}.

\textbf{Figure~\ref{fig:o-LiMnO2}b} depicts a map of the 0-TM capacity
as function of the Li content and the degree of disorder ranging from
the ideal ordered \ce{o-LiMnO2} structure (0\%~cation mixing) to the
fully random rocksalt structure (100\%~cation mixing).
As seen in the figure, the 0-TM percolation threshold decreases with
cation mixing from $\sim{}1.18$ for \ce{o-LiMnO2} to the rocksalt value
of $\sim{}1.09$.
In other words, around 20\% Li excess $x=1.2$ are needed for the
\ce{o-LiMnO2} structure to become 0-TM percolating.
On the other hand, the 0-TM capacity of compositions beyond $\sim{}25$\%
Li excess does not vary strongly with the degree of cation mixing.

The color-code in the map shown in \textbf{Fig.~\ref{fig:o-LiMnO2}c}
indicates the expected tortuosity for the same configuration and
composition space.
Note that the tortuosity increases with the degree of cation mixing.
Hence, while cation mixing lowers the percolation threshold in the case
of \ce{o-LiMnO2} it increases the tortuosity, and a degree of
\ce{o-LiMnO2}-like short-range order might therefore be beneficial for
the rate capability of the cathode material.

\section{Discussion}

\noindent
In this chapter, we reviewed the requirements for ionic conduction in
crystalline solids with substitutional disorder and introduced a lattice
model methodology for the simulation of properties related to the ionic
percolation behavior.
We furthermore introduced definitions for the percolation threshold, the
fraction of accessible sites, and the tortuosity for ionic conduction.

We applied the methodology to the example of Li-ion conduction in fully
and partially cation disordered rocksalt-type cathodes in which Li
transport is facilitated by fast 0-TM diffusion channels.
In the context of LIB cathode materials, the percolation threshold
determines whether a given composition is an ionic conductor, the
fraction of accessible sites can be converted to expected capacities,
and the tortuosity can be understood as a scaling factor for the
diffusivity and thus affects the ionic conductivity (rate capability).
However, the methodology is general and can be applied to any kind of
crystal structure and is agnostic of the diffusing species.

Some additional remarks regarding Li transport and the concept of 0-TM
diffusion are warranted:
First, in the discussed examples, we did not distinguish between Li and
vacancy sites despite mentioning that Li diffusion relies on a
di-vacancy mechanism.
This is rooted in the assumption that vacancies can always be created
along the percolating 0-TM diffusion pathway so that the diffusion
mechanism is, in fact, always the di-vacancy mechanism.
However, such a simplification is not always possible.
For example, in materials with a fixed vacancy concentration such as
solid electrolytes it might be required to model vacancies explicitly.
This is possible within the framework of ionic percolation theory, and
one example from the literature is the description of Mg percolation in
Mg transition metal spinels.\cite{cm29-2017-7918}

Second, at high vacancy concentrations some of the tetrahedral sites in
a rocksalt-type oxide may become more stable than the neighboring
octahedral sites, so that the diffusion mechanism locally changes from
\emph{octahedral--tetrahedral-octahedral} to
\emph{tetrahedral--octahedral--tetrahedral}.
This does not affect the overall Li diffusion pathway and therefore has
no effect on the percolation properties discussed in the present paper.
However, it is in general possible to distinguish between different
sublattices, such as octahedral and tetrahedral sites, in ionic
percolation simulations.\cite{cm29-2017-7918}

Third, it is important to keep in mind that the 0-TM capacity is a lower
bound on the expected capacity.
The activation energies of diffusion channels in disordered compounds
follow a distribution, and while nearly all 0-TM channels will have a
low barrier and will be active, some of the 1-TM channels might also
contribute to Li transport.
Hence, even compositions below the 0-TM percolation threshold will
typically exhibit some amount of capacity.

The tortuosity of diffusion pathways in disordered LIB cathodes had, to
the author's knowledge, not previously been considered.
Based on the herein made observations, in the composition space of most
cation-disordered cathode materials (\emph{i.e.}, compositions with Li
contents $1.10\leq{}x\leq{}1.20$ per \ce{Li_xM_{2-x}O2} formula unit),
the average tortuosity is with $\sim$1.2~to~$\sim$1.4 relatively high.
Such tortuosity corresponds to diffusion pathways that are 20\% to 40\%
longer than linear diffusion pathways, giving rise to an effective
reduction of the diffusivity by the same amount.
This observation is notable since the tortuosity of ordered cathode
materials in the layered, spinel, and \ce{o-LiMnO2} structures is equal
to~1, if 1-TM channels are active (a reasonable assumption in ordered
structures).
Short-range order can reduce the tortuosity of cation-disordered
materials and might therefore be beneficial for the ionic conductivity.

Finally, other factors not discussed in this chapter can affect the
performance of an ionic conductor.
In structures that are not percolating, the mean length of
(non-percolating) diffusion paths and the mean distance between
diffusion path domains can be useful criteria for estimating the ionic
conductivity.~\cite{nc10-2019-592}
The dimensionality of diffusion networks consisting of interconnected
diffusion paths can also play a role.
For example, one-dimensional diffusion pathways are more sensitive to
blockage by point defects,\cite{nl10-2010-4123} whereas cation disorder
typically creates three-dimensional diffusion networks.
Finally, the effect of realistic short-range order on the ionic
transport can be incorporated into the ionic percolation theory
framework, for example, in combination with MC sampling of the atomic
configurations.~\cite{nc10-2019-592, aem10-2020-1903240}

\section{Conclusions and final remarks}

\noindent
This book chapter reviewed percolation on regular lattices in the
context of ionic conductors and discussed applications to Li-ion
conduction in Li-ion battery cathode materials.
All examples are based on the free and open-source \emph{Dribble}
package, and Python notebooks with examples are available at
\url{https://github.com/atomisticnet/dribble}.
We showed how to estimate percolation thresholds, the fraction of
accessible sites, and the tortuosity of diffusion pathways for the
example of the disordered rocksalt structure and the related ordered
orthorhombic \ce{LiMnO2} structure.
We find that \ce{o-LiMnO2}-like short-range order in disordered rocksalt
cathodes is slightly detrimental for the percolation threshold,
increasing the required amount of Li excess from 9\% to 18\%.
At the same time, the short-range order reduces the tortuosity of the
diffusion pathways, which can be expected to be beneficial for the rate
capability.
The presented framework for percolation analysis is not limited to Li
transport or Li-ion battery materials, but can be generally applied to
the modeling of ionic conductors with substitutional disorder.


\raggedright
\bibliography{Urban}{}
\bibliographystyle{aipnum4-1}

\end{document}